\documentclass[fleqn,twoside,twocolumn,nofootinbib,showkeys]{revtex4} 
\usepackage[sec,nocpr]{ujp} 
\usepackage[cp1251]{inputenc}

\def\bA{{\boldsymbol A}}
\def\be{{\boldsymbol e}}
\def\bk{{\boldsymbol k}}
\def\bx{{\boldsymbol x}}

\def\cA{{\cal A}}
\def\cP{{\cal P}}

\def\ri{{\rm i}}

\def\rP{{\rm P}}

\begin{document}
\title[Inflationary Magnetogenesis with Helical Coupling]
{INFLATIONARY MAGNETOGENESIS\\ WITH HELICAL COUPLING\footnotemark[1]}%
\author{Yu.V.~Shtanov}
\affiliation{Bogolyubov Institute for Theoretical Physics, Nat.\@ Acad.\@ of Sci.\@ of Ukraine}
\address{14b, Metrologichna Str., Kiev 03143, Ukraine}
\email{shtanov@bitp.kiev.ua}
\affiliation{Astronomical Observatory, Taras Shevchenko National University of Kiev}%
\address{3, Observatorna Str., Kiev 04053, Ukraine}%
\author{M.V.~Pavliuk}
\affiliation{Department of Physics, Taras Shevchenko National University of Kiev}
\address{2, Academician Glushkov Ave., Kiev 03022, Ukraine}
\email{pavlyukconnection@gmail.com}

\udk{524.83}  \razd{\seci}

\autorcol{Yu.V.~Shtanov, M.V.~Pavliuk}

\setcounter{page}{1009}%

\begin{abstract}
We describe a simple scenario of inflationary magnetogenesis based on a helical coupling to electromagnetism.  It allows to generate helical magnetic fields of strength of order up to $10^{- 7}\,\text{G}$, when extrapolated to the current epoch, in a narrow spectral band centered at any physical wavenumber by adjusting the model parameters. Additional constraints on magnetic fields arise from the considerations of baryogenesis and, possibly, from the Schwinger effect of creation of charged particle-antiparticle pairs.  
\end{abstract}


\def\doi{https://doi.org/10.15407/ujpe64.11.1009}
\def\year{2019}
\def\volume{64}
\def\jnumber{11}

\maketitle

\setcounter{footnote}{1}

\section{Introduction}

\footnotetext[1]{This work is based on the results presented at the XI Bolyai--Gauss--Lobachevskii (BGL-2019) Conference: Non-Euclidean, Noncommutative Geometry and Quantum Physics}

Magnetic fields permeate our universe on various spatial scales \cite{KF}.  There is a strong indication of the presence of magnetic fields in intergalactic medium, including voids \cite{Tavecchio:2010mk, Ando:2010rb, Neronov:1900zz, Dolag:2010, Taylor:2011}, with strengths $B \gtrsim 10^{-16}\,\text{G} \times \text{max}\, \big\{ 1, \sqrt{\text{Mpc}/ \lambda} \big\}$, where $\lambda$ is the coherence length of the field.  These observations suggest a cosmological origin of magnetic fields, which are subsequently amplified in galaxies, presumably, by the dynamo mechanism (see reviews \cite{Durrer:2013pga, Subramanian:2015lua}).

Among the most attractive possibilities of generating magnetic fields is inflationary magnetogenesis; it naturally explains their large coherence length, which can be comparable to the size of the large-scale structure.  Amplification of the vacuum electromagnetic field during inflation requires violation of the conformal invariance of the field equations.  A simple suggestion \cite{Turner:1987bw, Ratra:1991bn} is to consider a modified gauge-invariant Lagrangian for the electromagnetic field of the form\footnote{In frames of the standard model, to preserve gauge invariance, one should consider coupling to the weak-hypercharge gauge field $B_\mu$, but we will not go into these subtleties here that do not modify our results.} 
\begin{equation}\label{Lem}
	{\cal L}_{\rm em} = - \frac14 I^2  F_{\mu\nu} F^{\mu\nu} - \frac14 f  F_{\mu\nu} \tilde F^{\mu\nu} \, ,
\end{equation}
where $\tilde F^{\mu\nu} = \frac12 \epsilon^{\mu\nu}{}_{\alpha\beta} F^{\alpha\beta}$ is the Hodge dual of $F^{\mu\nu}$, and $I$ and $f$ are non-trivial functions of time on the stage of inflation due to their dependence on the background fields such as the inflaton or the metric curvature.  The first term in (\ref{Lem}) is the so-called kinetic coupling; the second, parity-violating term, is the helical coupling.  Numerous versions of this model have been under consideration in the literature (see \cite{Durrer:2013pga, Subramanian:2015lua} for recent reviews). 

Scenarios with the kinetic coupling to electromagnetism face the problems of back-reaction and strong gauge coupling \cite{Demozzi:2009fu, Urban:2011bu, BazrafshanMoghaddam:2017zgx}.  Essentially, if the function $I$ is monotonically decreasing with time, then it is electric field that is predominantly enhanced, causing the problem of back-reaction on inflation and preventing generation of magnetic fields of plausible strengths.  If the function $I$ is monotonically increasing, then magnetic field is enhanced predominantly, but there arises the problem of strong gauge coupling invalidating the perturbative approach because the effective gauge coupling $e_{\rm eff} \propto I^{-1}$ is too large at the early stages of this process \cite{Demozzi:2009fu}.  Typical attempts at overcoming these difficulties \cite{Caprini:2014mja, Sharma:2017eps, Sharma:2018kgs} require sufficiently low scale of inflation to produce magnetic fields of considerable magnitudes.

In this paper, we consider model (\ref{Lem}) with the standard kinetic term ($I \equiv 1$)  but with a non-trivial helical-coupling function $f$.  Contrary to the case of kinetic coupling, the absolute value of $f$ is of no significance (since the second term in (\ref{Lem}) with constant $f$ is topological), which greatly broadens the scope of its possible evolution\,---\,the strong-coupling problem does not arise in this model.  Typical laws of evolution of $f$ were previously studied in the literature.   Evolution in the form $f \propto (\log a)^s$ during inflation (arising in the case of linear dependence of $f$ on the inflaton field) generically leads to maximally helical magnetic fields with blue power spectrum \cite{Durrer:2010mq, Jain:2012jy, Fujita:2015iga} and with too little power on the comoving scales of galaxies, clusters and voids to account for the magnetic fields in these objects \cite{Durrer:2013pga}.  Evolution in the form of power law $f \propto a^s$ with $s > 0$ \cite{Campanelli:2008kh} also allows for only a negligible amplification of magnetic field \cite{Shtanov:2019civ}.

In this paper, we describe a different scenario, in which $f$ evolves monotonically during some period of time, interpolating between two asymptotic constant values.  By adjusting the two free parameters of this model (the duration of the evolution period and the change $\Delta f$ during this period), helical magnetic fields of strength of order up to $10^{- 7}\,\text{G}$ at the current epoch can be generated in a narrow spectral band that can be centered at any reasonable wavenumber.

Within the standard model of electroweak interactions, magnetogenesis proceeds as follows.  Depending on whether or not the electroweak symmetry is broken during inflation, 
either electromagnetic or weak-hypercharge field is generated. 
After reheating, the electroweak symmetry is restored, and only the weak-hypercharge part of the magnetic field survives, making the field hypermagnetic.  This field evolves till the electroweak crossover at temperature about 160~GeV, during which it is gradually transformed to the usual magnetic field that survives until the present epoch \cite{Kajantie:1996qd, DOnofrio:2015gop}. 

Helical hypermagnetic fields source the baryon number in the hot universe \cite{Giovannini:1997gp, Giovannini:1997eg}, opening up an interesting possibility of baryogenesis \cite{Bamba:2006km, Anber:2015yca, Fujita:2016igl, Kamada:2016eeb, Kamada:2016cnb, Jimenez:2017cdr}.  The requirement of not exceeding the observed baryon number density then imposes additional constraints on the strength and coherence length of maximally helical primordial magnetic field.  


\section{Scenario}
\label{sec:scenario}

We consider the model with Lagrangian (\ref{Lem}), in which $I \equiv 1$, and $f$ is a function of time through its dependence on the background fields (such as the inflaton and/or the metric curvature).

We work with a spatially flat metric in conformal coordinates, $ds^2=a^2 (\eta) \left( d\eta^2 - \delta_{ij} d x^i d x^j \right)$, and adopt the longitudinal gauge $A_0 = 0$ and $\partial^i A_i = 0$ for the vector potential. 
In the spatial Fourier representation, and in the constant normalized helicity basis $\be_h (\bk)$, $h = \pm 1$, such that $\ri \bk \times \be_h = h k \be_h$, we have $\bA = \sum_h \cA_h \be_h e^{\ri \bk \bx}$.  The helicity components then satisfy the equation
\begin{equation} \label{eqAh}	
	\cA''_h + \left( k^2 + h k f' \right) \cA_h = 0 \, , 
\end{equation}
where the prime denotes the derivative with respect to the conformal time $\eta$.

The spectral densities of quantum fluctuations of magnetic and electric field are characterized, respectively, by the usual relations\footnote{We are using units in which $\hbar = c = 1$.}
\begin{align}
	\cP_B &= \frac{d \rho_B}{d \ln k} = \frac{k^4}{8 \pi^2 a^4} \sum_h \left| \cA_h (\eta, k) \right|^2 \, , \label{specB} \\
	\cP_E &= \frac{d \rho_E}{d \ln k} = \frac{k^4}{8 \pi^2 a^4} \sum_h \left| \frac{\cA'_h (\eta, k)}{k} \right|^2 \, , \label{specE}
\end{align}
in which the amplitude of the vector potential is normalized so that $\cA_h (\eta) \sim e^{- \ri k \eta}$ as $\eta \to - \infty$.  The factor in front of the sums in (\ref{specB}) and (\ref{specE}) is the spectral density of vacuum fluctuations in flat space-time in each mode at the physical wavenumber $k/a$.

For sufficiently low values of $k$, the term $h k f'$ in the brackets of equation (\ref{eqAh}) dominates over $k^2$, and, for the helicity for which this term has negative sign, one  expects a regular growth of the corresponding mode.  

Basing on this observation, we considered in \cite{Shtanov:2019civ} a linear dependence of $f$ as a function of conformal time: $f' (\eta) = \text{const}$, $\eta_1 \leq \eta \leq \eta_2$, joining the time epochs with constant values of $f$.  The growth of $\cA_h$ in this case is exponential for one of the helicities in the interval $\eta_1 < \eta < \eta_2$.  In the present paper, we consider a qualitatively similar but smooth evolution of $f (\eta)$, for which the problem is also exactly solvable.  By conveniently shifting the origin of the cosmological time $\eta$, we take the coupling function in the form
\begin{equation}\label{f}
	f (\eta) = \frac12 \Delta f \tanh \left( \frac{2 \eta}{\Delta \eta} \right) \, ,
\end{equation}
in which $\Delta \eta > 0$ is its temporal width.  Then $f'_0 \equiv f'(0) = \Delta f / \Delta \eta$.  We assume that the evolution of $f$ essentially completes by the end of inflation.

Introducing the dimensionless time $\tau = 2 \eta / \Delta \eta$ and wavenumber $p = k \Delta \eta / 2$, one can write the general solution of (\ref{eqAh}) in terms of the Ferrers functions \cite[Chapter~14]{Olver}:
\begin{equation} \label{solA}
	\cA_h ( \tau ) = c_+ \rP_\nu^\mu \left( \tanh \tau \right) + c_- \rP_\nu^{-\mu} \left( \tanh \tau \right) \, ,	
\end{equation}
with
\begin{equation} \label{munu} 
	\mu = \ri p \, , \quad \nu = q - \frac12  \, , \quad q \equiv \frac12 \sqrt{1 + 2 h \Delta f p} \, . 
\end{equation}

The asymptotics $\cA_h \sim e^{- \ri k \eta} = e^{- \ri p \tau}$ as $ \tau \to - \infty$ determines the constants $c_+$ and $c_-$ in (\ref{solA}).  By considering the opposite asymptotics $\cA_h \sim \alpha_k e^{- \ri p \tau} + \beta_k e^{\ri p \tau} $ as $ \tau \to \infty$ in (\ref{solA}), one reads off the Bogolyubov's coefficients $\alpha_k$ and $\beta_k$.  Skipping a simple calculation, we present here the result: 
\begin{align}
	\alpha_k &= \frac{\Gamma (1 - \ri p) \Gamma (- \ri p)}{\Gamma \left( \frac12 + q - \ri p \right) \Gamma \left( \frac12 - q - \ri p \right)} \, , \\
	\beta_k &= - \ri \frac{\cos ( \pi q )}{\sinh (\pi p ) } \, ,
\end{align}
with the required property $\left| \alpha_k \right|^2 - \left| \beta_k \right|^2 = 1$.  
After the evolution of the coupling, the mean number of quanta in a given mode is 
\begin{equation}\label{nk}
n_k = \left| \beta_k \right|^2 = \frac{\cos^2 (\pi q )}{\sinh^2 (\pi p )} \, .
\end{equation}

For the helicity satisfying $h \Delta f < 0$, the quantity $q$, given by (\ref{munu}), is purely imaginary for $p > 1 / 2 |\Delta f|$.  In the approximation 
$p \gg \text{max}\,\left\{ {1}/{2 |\Delta f|}\, , \, {1}/{\pi} \right\}$,
we then obtain
\begin{equation} \label{n}
	n_k \approx e^{2 \pi \left( \sqrt{|\Delta f| p / 2} - p \right)} = e^{\pi | \Delta f | \left( \sqrt{ k / k_0} - k / k_0 \right) } \, ,
\end{equation}
where $k_0 = \left| f_0' \right|$ is the wavenumber at which spectrum (\ref{n}) reaches unity during the exponential decline.  The exponent of this expression reaches maximum at $k_{\rm m} = k_0 / 4$, with the maximum mean occupation number $n_{\rm m} = e^{\pi | \Delta f| / 4}$, which is exponentially large for $| \Delta f | \gg 1$. 
Spectrum (\ref{nk}) is plotted in Fig.~\ref{fig:plot} on a logarithmic scale for a typical value $|\Delta f| = 250$ (see below).  The mean occupation numbers for the opposite helicity can be neglected. 

\begin{figure}
	\includegraphics[width=\column]{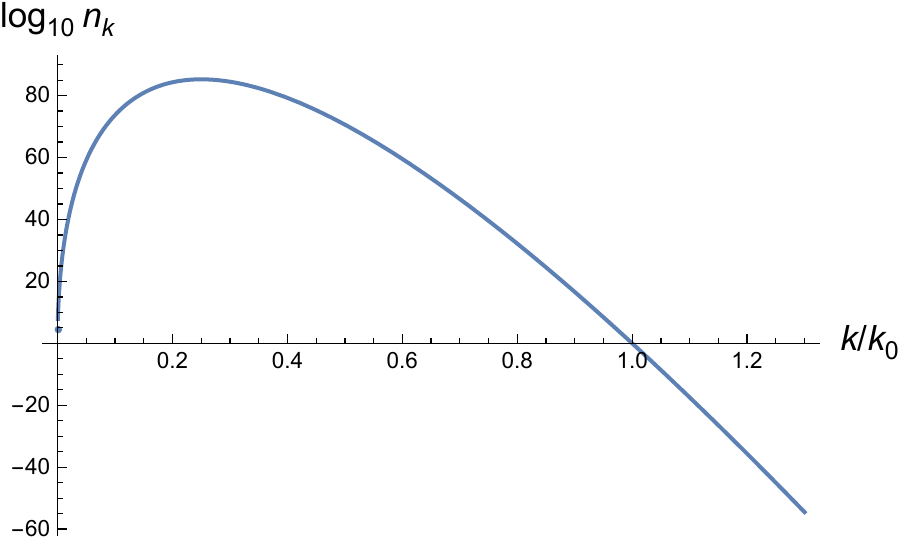}
	\vskip-3mm\caption{Spectrum (\ref{nk}) on a logarithmic scale for the helicity satisfying $h \Delta f < 0$ and for $|\Delta f| = 250$. \label{fig:plot}}
\end{figure}

The spectral densities (\ref{specB}) and (\ref{specE}) are of comparable magnitudes, and,  
since one of the helicities is dominating in $n_k$, we have, using (\ref{n}), 
\begin{equation}\label{PBE}
	\cP_B \approx \cP_E \approx \frac{k^4}{4 \pi^2 a^4} \left[ 1 + e^{\pi | \Delta f | \left( \sqrt{ k / k_0} - k / k_0 \right) } \right] \, .
\end{equation}

We observe that the spectral densities are peaked at the central value $k = k_{\rm m}$ with width $\Delta k \simeq k_{\rm m} / \sqrt{|\Delta f|} \ll k_{\rm m}$ for $|\Delta f| \gg 1$.  Thus, electric and magnetic fields are generated in this scenario with similar spectra in the spectral region of amplification.

Using a Gaussian approximation to (\ref{PBE}), one can estimate the total excess (over vacuum) of the electromagnetic energy density and the magnetic field:
\begin{align} \label{em}
	\rho_{\rm em} &\simeq \frac{2 k_{\rm m}^4}{\pi^2 a^{4}} \frac{1}{| \Delta f|^{1/2}} e^{\pi | \Delta f |/4} \, , \\
	B &\simeq \sqrt{\rho_{\rm m}} \simeq \frac{\sqrt{2} k_{\rm m}^2}{\pi a^{2}} \frac{1}{| \Delta f|^{1/4}} e^{\pi | \Delta f |/8} \, . \label{main}
\end{align}

Expressions (\ref{PBE})--(\ref{main}) contain two free parameters of the theory, $\Delta f$ and $k_{\rm m} = k_0/4 = | \Delta f | / 4 \Delta \eta$, which can be adjusted to produce magnetic fields of desirable strength with spectral density centered at the wavenumber $k_{\rm m}$.  For instance, in order to obtain $B_0 \simeq 3 \times 10^{-16}\, \text{G}$ at the current epoch with spectrum peaked on the comoving scale $k_{\rm m}/a_0 \simeq \text{Mpc}^{-1}$, we require $|\Delta f| \approx 250$.  
The dependence of $|\Delta f|$ on the magnitude $B_0$ is quite weak (logarithmic); thus, for $B_0$ in the range $10^{- 30}$$-$$10^{- 7}$\,G on the same scale, one requires $|\Delta f| \simeq 165$$-$$300$.  

\section{Back-reaction on inflation and baryogenesis}
\label{sec:inflation}

Back-reaction on inflation and on the dynamics of the inflaton field in the scenario under consideration was estimated in \cite{Shtanov:2019civ} with the upper bound 
\begin{equation} 
	B_0 \ll 3 \times 10^{-7} g_r^{-1/6} \, \text{G}
\end{equation}
on the magnetic field extrapolated today, assuming that its generation was completed right by the end of inflation. Here, $g_r$ is the number of relativistic degrees of freedom in the universe after reheating. 

One of the most interesting effects of evolving helical hypermagnetic fields is generation of baryon number in the early hot universe \cite{Giovannini:1997gp, Giovannini:1997eg}. This opens up an intriguing possibility of explaining the observed baryon asymmetry ($\eta_{\rm b} = n_{\rm b} / s \sim 10^{-10}$, where $n_{\rm b}$ is the baryon number density, and $s$ is the entropy density in the late-time universe) \cite{Bamba:2006km, Anber:2015yca, Fujita:2016igl, Kamada:2016eeb, Kamada:2016cnb, Jimenez:2017cdr}.  
According to the recent estimates \cite{Kamada:2016cnb}, the resulting baryon asymmetry, when expressed through the current strength $B_0$ and coherence length $\lambda_0$ of (originally maximally helical) magnetic field, turns out to be (with some theoretical uncertainty in the exponent)
\begin{equation}\label{basym}
	\eta_{\rm b} \sim 10^{- (9 \text{--} 12)} \frac{\lambda_0}{\text{Mpc}} \left( \frac{B_0}{10^{-21}\, \text{G}} \right)^2 \, .
\end{equation} 
One can see that the present model of magnetogenesis can also support baryogenesis.  On the other hand, as follows from (\ref{basym}), to avoid overproduction of the baryon number, a model of magnetogenesis should respect a constraint on the current strength and coherence length of magnetic field, provided it was originally maximally helical and existed prior to the electroweak crossover:
\begin{equation}\label{bacon}
	B_0 \lesssim 10^{-21}\, \left( \frac{\text{Mpc}}{\lambda_0} \right)^{1/2} \, \text{G} \, .
\end{equation}
With $\lambda_0 \sim a_0/k_{\rm m}$, this constrains the possible values of $B_0$ and $k_{\rm m}$ in the present scenario.
	

\section{Summary}
\label{sec:summary}

We proposed a simple model of inflationary magnetogenesis based on the helical coupling in Lagrangian (\ref{Lem}).  In our case, the coupling $f$ evolves monotonically during a finite time interval, interpolating between two constant values in the past and in the future.  The duration of the transition $\Delta \eta$ and the corresponding change $\Delta f$ are the two parameters of the model that can be adjusted to produce magnetic field of any strength in a narrow spectral band centered at any reasonable comoving wavenumber $k_{\rm m} = |\Delta f| / 4 \Delta \eta$.  For the simple inflation based on a massive scalar field, this scenario allows for production of magnetic fields with extrapolated current values up to $B_0 \sim 10^{- 7}\, \text{G}$. 

Primordial helical hypermagnetic fields may be responsible for generating baryon asymmetry of the universe \cite{Bamba:2006km, Anber:2015yca, Fujita:2016igl, Kamada:2016eeb, Kamada:2016cnb, Jimenez:2017cdr}.  This imposes a post-inflationary constraint (\ref{bacon}) on the admissible values of $B_0$ and $k_{\rm m}$ in our simple scenario of monotonic evolution of $f$.  Other constraints on models of this type may arise from the considerations of the created baryon number inhomogeneities \cite{Giovannini:1997gp, Giovannini:1997eg} that can affect the cosmic microwave background and primordial nucleosynthesis.  This problem, specific to the discussed baryogenesis scenario, requires special investigation.  Another important issue that awaits for future analysis in the present scenario is the Schwinger effect of creation of charged particle-antiparticle pairs during magnetogenesis \cite{Sharma:2017eps, Kobayashi:2014zza, Sobol:2018djj, Sobol:2019xls}. 

\vskip3mm \textit{This work was supported by the National Academy of Sciences of Ukraine (project 0116U003191) and by the scientific program ``Astronomy and Space Physics'' (project 19BF023-01) of the Taras Shevchenko National University of Kiev.}

\vspace*{-5mm} \rezume{%
Ю.В.~Штанов, М.В.~Павлюк} {ІНФЛЯЦІЙНИЙ МАГНІТОГЕНЕЗ З ГЕЛІКАЛЬНИМ ЗВ'ЯЗКОМ} {Описано простий сценарій інфляційного магнітогенезу, оснований на гелікальному зв'язку з електромагнетизмом.  Він дозволяє генерувати гелікальні магнітні поля з напруженістю до $10^{- 7}\,\text{Гс}$ у сучасну епоху у вузькій спектральній смузі, центрованій на довільному фізичному хвильовому числі, через налаштування параметрів моделі.  Додаткові обмеження на магнітне поле виникають із міркувань теорії баріогенезу та, ймовірно, з ефекту Швінгера народження заряджених пар частинок-античастинок.}

\end{document}